\documentclass[pre,floatfix,twocolumn]{revtex4}

\usepackage{amsmath}
\usepackage{amsfonts}
\usepackage{mathrsfs}
\usepackage{epsfig}
\usepackage{graphicx}
\usepackage{dcolumn}
\usepackage{doi}

\newcommand{\br}[1]{[#1]}
\newcommand{\dt}[1]{\frac{d #1}{dt}}

\makeatletter
\newcommand{\FF}{\afterassignment\FF@aux\count0=}
\newcommand{\FF@aux}{\csname FF\the\count0\endcsname}
\makeatother

\makeatletter
\newcommand{\F}{\afterassignment\F@aux\count0=}
\newcommand{\F@aux}{\csname F\the\count0\endcsname}
\makeatother

\makeatletter
\newcommand{\GG}{\afterassignment\GG@aux\count0=}
\newcommand{\GG@aux}{\csname GG\the\count0\endcsname}
\makeatother

\makeatletter
\newcommand{\G}{\afterassignment\G@aux\count0=}
\newcommand{\G@aux}{\csname G\the\count0\endcsname}
\makeatother

\makeatletter
\newcommand{\LL}{\afterassignment\LL@aux\count0=}
\newcommand{\LL@aux}{\csname LL\the\count0\endcsname}
\makeatother

\makeatletter
\newcommand{\LM}{\afterassignment\LM@aux\count0=}
\newcommand{\LM@aux}{\csname LM\the\count0\endcsname}
\makeatother

\makeatletter
\newcommand{\FG}{\afterassignment\FG@aux\count0=}
\newcommand{\FG@aux}{\csname FG\the\count0\endcsname}
\makeatother

\makeatletter
\newcommand{\LD}{\afterassignment\LD@aux\count0=}
\newcommand{\LD@aux}{\csname LD\the\count0\endcsname}
\makeatother

\expandafter\newcommand\csname FF1\endcsname{[3K]}
\expandafter\newcommand\csname FF2\endcsname{[S.3K]}
\expandafter\newcommand\csname FF3\endcsname{[3K^{*}.P_{3K}]}
\expandafter\newcommand\csname FF4\endcsname{[3K^{*}]}
\expandafter\newcommand\csname FF5\endcsname{[2K]}
\expandafter\newcommand\csname FF6\endcsname{[3K^{*}.2K]}
\expandafter\newcommand\csname FF7\endcsname{[2K^{*}.P_{2K}]}
\expandafter\newcommand\csname FF8\endcsname{[2K^{*}]}
\expandafter\newcommand\csname FF9\endcsname{[3K^{*}.2K^{*}]}
\expandafter\newcommand\csname FF10\endcsname{[2K^{**}.P_{2K}]}
\expandafter\newcommand\csname FF11\endcsname{[2K^{**}]}
\expandafter\newcommand\csname FF12\endcsname{[K]}
\expandafter\newcommand\csname FF13\endcsname{[2K^{**}.K]}
\expandafter\newcommand\csname FF14\endcsname{[K^{*}.P_{K}]}
\expandafter\newcommand\csname FF15\endcsname{[K^{*}]}
\expandafter\newcommand\csname FF16\endcsname{[2K^{**}.K^{*}]}
\expandafter\newcommand\csname FF17\endcsname{[K^{**}.P_{K}]}
\expandafter\newcommand\csname FF18\endcsname{[K^{**}]}

\expandafter\newcommand\csname GG1\endcsname{[P^{f}_{3K}]}
\expandafter\newcommand\csname GG2\endcsname{[P^{f}_{2K}]}
\expandafter\newcommand\csname GG3\endcsname{[P^{f}_{2K}]}
\expandafter\newcommand\csname GG4\endcsname{[P^{f}_{K}]}
\expandafter\newcommand\csname GG5\endcsname{[P^{f}_{K}]}

\expandafter\newcommand\csname F1\endcsname{F_{1}}
\expandafter\newcommand\csname F2\endcsname{F_{2}}
\expandafter\newcommand\csname F3\endcsname{F_{3}}
\expandafter\newcommand\csname F4\endcsname{F_{4}}
\expandafter\newcommand\csname F5\endcsname{F_{5}}
\expandafter\newcommand\csname F6\endcsname{F_{6}}
\expandafter\newcommand\csname F7\endcsname{F_{7}}
\expandafter\newcommand\csname F8\endcsname{F_{8}}
\expandafter\newcommand\csname F9\endcsname{F_{9}}
\expandafter\newcommand\csname F10\endcsname{F_{10}}
\expandafter\newcommand\csname F11\endcsname{F_{11}}
\expandafter\newcommand\csname F12\endcsname{F_{12}}
\expandafter\newcommand\csname F13\endcsname{F_{13}}
\expandafter\newcommand\csname F14\endcsname{F_{14}}
\expandafter\newcommand\csname F15\endcsname{F_{15}}
\expandafter\newcommand\csname F16\endcsname{F_{16}}
\expandafter\newcommand\csname F17\endcsname{F_{17}}
\expandafter\newcommand\csname F18\endcsname{F_{18}}

\expandafter\newcommand\csname G1\endcsname{G_{1}}
\expandafter\newcommand\csname G2\endcsname{G_{2}}
\expandafter\newcommand\csname G3\endcsname{G_{3}}
\expandafter\newcommand\csname G4\endcsname{G_{4}}
\expandafter\newcommand\csname G5\endcsname{G_{5}}

\expandafter\newcommand\csname LL1\endcsname{k_{-1}}
\expandafter\newcommand\csname LL2\endcsname{kp_{2}}
\expandafter\newcommand\csname LL3\endcsname{k_{1}}
\expandafter\newcommand\csname LL4\endcsname{k_{2}}
\expandafter\newcommand\csname LL5\endcsname{kp_{1}}
\expandafter\newcommand\csname LL6\endcsname{kp_{-1}}
\expandafter\newcommand\csname LL7\endcsname{k_{-3}}
\expandafter\newcommand\csname LL8\endcsname{k_{4}}
\expandafter\newcommand\csname LL9\endcsname{k_{3}}
\expandafter\newcommand\csname LL10\endcsname{k_{-5}}
\expandafter\newcommand\csname LL11\endcsname{k_{6}}
\expandafter\newcommand\csname LL12\endcsname{k_{5}}
\expandafter\newcommand\csname LL19\endcsname{kp_{4}}
\expandafter\newcommand\csname LL20\endcsname{kp_{3}}
\expandafter\newcommand\csname LL21\endcsname{kp_{-3}}
\expandafter\newcommand\csname LL22\endcsname{kp_{5}}
\expandafter\newcommand\csname LL23\endcsname{kp_{6}}
\expandafter\newcommand\csname LL24\endcsname{kp_{-5}}
\expandafter\newcommand\csname LL25\endcsname{k_{-7}}
\expandafter\newcommand\csname LL26\endcsname{k_{8}}
\expandafter\newcommand\csname LL27\endcsname{k_{7}}
\expandafter\newcommand\csname LL28\endcsname{k_{-9}}
\expandafter\newcommand\csname LL29\endcsname{k_{10}}
\expandafter\newcommand\csname LL30\endcsname{k_{9}}
\expandafter\newcommand\csname LL31\endcsname{kp_{8}}
\expandafter\newcommand\csname LL32\endcsname{kp_{7}}
\expandafter\newcommand\csname LL33\endcsname{kp_{-7}}
\expandafter\newcommand\csname LL34\endcsname{kp_{10}}
\expandafter\newcommand\csname LL35\endcsname{kp_{9}}
\expandafter\newcommand\csname LL36\endcsname{kp_{-9}}

\expandafter\newcommand\csname LM1\endcsname{\lambda_{1}}
\expandafter\newcommand\csname LM2\endcsname{\lambda_{2}}
\expandafter\newcommand\csname LM3\endcsname{\lambda_{3}}
\expandafter\newcommand\csname LM4\endcsname{\lambda_{4}}
\expandafter\newcommand\csname LM5\endcsname{\lambda_{5}}
\expandafter\newcommand\csname LM6\endcsname{\lambda_{6}}
\expandafter\newcommand\csname LM7\endcsname{\lambda_{7}}
\expandafter\newcommand\csname LM8\endcsname{\lambda_{8}}
\expandafter\newcommand\csname LM9\endcsname{\lambda_{9}}
\expandafter\newcommand\csname LM10\endcsname{\lambda_{10}}
\expandafter\newcommand\csname LM11\endcsname{\lambda_{11}}
\expandafter\newcommand\csname LM12\endcsname{\lambda_{12}}
\expandafter\newcommand\csname LM13\endcsname{\lambda_{13}}
\expandafter\newcommand\csname LM14\endcsname{\lambda_{14}}
\expandafter\newcommand\csname LM15\endcsname{\lambda_{15}}
\expandafter\newcommand\csname LM16\endcsname{\lambda_{16}}
\expandafter\newcommand\csname LM17\endcsname{\lambda_{17}}
\expandafter\newcommand\csname LM18\endcsname{\lambda_{18}}
\expandafter\newcommand\csname LM19\endcsname{\lambda_{19}}
\expandafter\newcommand\csname LM20\endcsname{\lambda_{20}}
\expandafter\newcommand\csname LM21\endcsname{\lambda_{21}}
\expandafter\newcommand\csname LM22\endcsname{\lambda_{22}}
\expandafter\newcommand\csname LM23\endcsname{\lambda_{23}}
\expandafter\newcommand\csname LM24\endcsname{\lambda_{24}}
\expandafter\newcommand\csname LM25\endcsname{\lambda_{25}}
\expandafter\newcommand\csname LM26\endcsname{\lambda_{26}}
\expandafter\newcommand\csname LM27\endcsname{\lambda_{27}}
\expandafter\newcommand\csname LM28\endcsname{\lambda_{28}}
\expandafter\newcommand\csname LM29\endcsname{\lambda_{29}}
\expandafter\newcommand\csname LM30\endcsname{\lambda_{30}}
\expandafter\newcommand\csname LM31\endcsname{\lambda_{31}}
\expandafter\newcommand\csname LM32\endcsname{\lambda_{32}}
\expandafter\newcommand\csname LM33\endcsname{\lambda_{33}}
\expandafter\newcommand\csname LM34\endcsname{\lambda_{34}}
\expandafter\newcommand\csname LM35\endcsname{\lambda_{35}}
\expandafter\newcommand\csname LM36\endcsname{\lambda_{36}}
\expandafter\newcommand\csname LM37\endcsname{\lambda_{37}}
\expandafter\newcommand\csname LM38\endcsname{\lambda_{38}}
\expandafter\newcommand\csname LM39\endcsname{\lambda_{39}}
\expandafter\newcommand\csname LM40\endcsname{\lambda_{40}}
\expandafter\newcommand\csname LM41\endcsname{\lambda_{41}}
\expandafter\newcommand\csname LM42\endcsname{\lambda_{42}}
\expandafter\newcommand\csname LM43\endcsname{\lambda_{43}}
\expandafter\newcommand\csname LM44\endcsname{\lambda_{44}}
\expandafter\newcommand\csname LM45\endcsname{\lambda_{45}}
\expandafter\newcommand\csname LM46\endcsname{\lambda_{46}}
\expandafter\newcommand\csname LM47\endcsname{\lambda_{47}}
\expandafter\newcommand\csname LM48\endcsname{\lambda_{48}}
\expandafter\newcommand\csname LM49\endcsname{\lambda_{49}}
\expandafter\newcommand\csname LM50\endcsname{\lambda_{50}}
\expandafter\newcommand\csname LM51\endcsname{\lambda_{51}}
\expandafter\newcommand\csname LM52\endcsname{\lambda_{52}}
\expandafter\newcommand\csname LM53\endcsname{\lambda_{53}}
\expandafter\newcommand\csname LM54\endcsname{\lambda_{54}}

\expandafter\newcommand\csname FG1\endcsname{\gamma_{1}\,F_{1}^{\dagger}}
\expandafter\newcommand\csname FG2\endcsname{\gamma_{2}\,F_{2}^{\dagger}}
\expandafter\newcommand\csname FG3\endcsname{\gamma_{3}\,F_{3}^{\dagger}}
\expandafter\newcommand\csname FG4\endcsname{\gamma_{4}\,F_{4}^{\dagger}}
\expandafter\newcommand\csname FG5\endcsname{\gamma_{5}\,F_{5}^{\dagger}}
\expandafter\newcommand\csname FG6\endcsname{\gamma_{6}\,F_{6}^{\dagger}}
\expandafter\newcommand\csname FG7\endcsname{\gamma_{7}\,F_{7}^{\dagger}}
\expandafter\newcommand\csname FG8\endcsname{\gamma_{8}\,F_{8}^{\dagger}}
\expandafter\newcommand\csname FG9\endcsname{\gamma_{9}\,F_{9}^{\dagger}}
\expandafter\newcommand\csname FG10\endcsname{\gamma_{10}\,F_{10}^{\dagger}}
\expandafter\newcommand\csname FG11\endcsname{\gamma_{11}\,F_{11}^{\dagger}}
\expandafter\newcommand\csname FG12\endcsname{\gamma_{12}\,F_{12}^{\dagger}}
\expandafter\newcommand\csname FG13\endcsname{\gamma_{13}\,F_{13}^{\dagger}}
\expandafter\newcommand\csname FG14\endcsname{\gamma_{14}\,F_{14}^{\dagger}}
\expandafter\newcommand\csname FG15\endcsname{\gamma_{15}\,F_{15}^{\dagger}}
\expandafter\newcommand\csname FG16\endcsname{\gamma_{16}\,F_{16}^{\dagger}}
\expandafter\newcommand\csname FG17\endcsname{\gamma_{17}\,F_{17}^{\dagger}}
\expandafter\newcommand\csname FG18\endcsname{\gamma_{18}\,F_{18}^{\dagger}}

\begin{document}
\title{Non-associative learning in intra-cellular signaling networks}

\author{Tanmay Mitra$^{1,2}$, Shakti N. Menon$^1$ and Sitabhra Sinha$^{1,2}$}
\affiliation{$^1$The Institute of Mathematical Sciences, CIT Campus,
Taramani, Chennai 600113, India.\\
$^2$Homi Bhabha National Institute, Anushaktinagar, Mumbai 400094, India.}
\date{\today}


\begin{abstract}
Nonlinear systems driven by recurrent signals are known to exhibit
complex
dynamical responses which, in the physiological context, can have
important functional consequences.
One of the simplest biological systems that is exposed to
such repeated stimuli is the intra-cellular signaling network.
In this paper we investigate the periodic activation of an
evolutionarily
conserved motif of this network, viz., the
mitogen-activated protein kinase (MAPK) signaling cascade, with a
train of pulses.
The resulting response of the cascade, which shows integrative
capability over several successive pulses, is characterized by complex
adaptive
behavior. These include aspects of non-associative learning, in
particular,
habituation and sensitization, which are observed in response to high-
and
low-frequency stimulation, respectively.
In addition, the existence of a response threshold of the cascade,
an apparent refractory behavior following stimulation with short
inter-pulse interval,
and an alternans-like response under certain conditions suggest an
analogy with excitable media.
\end{abstract}


\maketitle



Nonlinear systems can respond to variations in their environment by 
exhibiting a wide range of complex dynamical
patterns~\cite{Testa1982,Glass1987,Cross1993,Lin2000,Pikovsky2003,Rabinovich2006} that 
may often be functionally
significant~\cite{Aschoff1981,Glass2001,Nelson2004,Stavreva2009,Heinrich2012,Lahav2013,Albeck2013}.
These variations are commonly associated with natural cycles such as
the diurnal rhythm. In particular, biological systems are typically
subjected to periodic stimuli with frequencies that can vary over a
wide range of time scales, viz., from ultradian to infradian
rhythms~\cite{Cauter1990,Lloyd1991,Schibler2005}. Examples
include the entrainment of the circadian clock to the day-night
cycle~\cite{Wright2013}, variations in hormonal levels over a period
of a month that drive
the menstrual cycle~\cite{Sherman1975} and calcium oscillations at the
time-scale of minutes
modulating the efficiency and specificity of gene
expression~\cite{Dolmetsch1998}. 
Of all the biological systems capable of exhibiting complex
functionally significant 
responses when driven by periodic stimuli, perhaps one of the simplest
is the intra-cellular signaling network~\cite{Li1992}.
In its natural environment, the membrane-bound receptors of a cell may
repeatedly be stimulated on encountering ligands, 
for instance as a consequence of pulsatile variations in
hormones~\cite{Leng1988}.
Cellular functions may also be
modulated by internal cues that vary periodically, e.g., oscillations
in the concentrations of intracellular messengers such as
Ca$^{2+}$~\cite{Berridge1990,Tsien1990} and cyclic
AMP~\cite{Dyachok2006,Borodinsky2006}. It is therefore important to
investigate how key components of the signaling network in the cell 
respond when subjected to repeated stimuli in the form of pulse trains.

One of the most ubiquitous motifs of this network is the
mitogen-activated protein kinase (MAPK) cascade, which is found across
all eukaryotic cells~\cite{Johnson1999,Seger1995}. It consists of a
sequential arrangement of three types of protein kinase, viz., MAPK,
MAPK kinase (MAP2K) and MAPK kinase kinase (MAP3K). The activated
kinase in each layer of the cascade functions as an enzyme for
phosphorylating (and thus activating) the kinase in the layer
immediately downstream. The subsequent deactivation is mediated by the
corresponding dephosphorylating enzyme known as phosphatases (P'ase).
Despite its structural simplicity this motif is involved in regulating
a wide array of vital cellular functions, including proliferation and
apoptosis~\cite{Seger1995}, stress response~\cite{Cooper1995} and gene
expression~\cite{Karin1996}. Activation of the cascade is initiated
when extracellular ligands stimulate membrane-bound receptors, or when
intracellular cues occur upstream of the cascade, with the information
being relayed to MAP3K through a series of intermediaries. The terminal
kinase of the motif (MAPK), transmits the signal further downstream 
by phosphorylating various proteins including transcription
regulators~\cite{Alberts6thed}. The behavior of the cascade subjected
to sustained stimulation has been extensively investigated, and the
existence of several emergent features has been observed. These include
ultrasensitivity~\cite{Ferrell1996}, bistability which allows the
system to switch between two states corresponding to low and high
activity~\cite{Ferrell2002,Kholodenko2004,KholodenkoBS2006,Herzel2007,Qiao2007}
and
oscillations~\cite{Qiao2007,Kholodenko2000,Ventura2008,Sontag2008,Shankaran2009,Kochanczyk2017}.
In earlier work, we have shown that the cascade stimulated with a pulse
of finite duration responds with a rich variety of transient behavior,
including phenomena indicative of the presence of short-term
memory~\cite{Mitra2018}. The complex modulations seen in the response
of the cascade are crucially dependent on the interactions between the
time-scales of the intrinsic processes and that of the applied
stimulus. It is thus intriguing to consider how the system will respond
to repeated stimulation.

In this paper, we investigate the dynamics of the MAPK cascade that is
stimulated by periodic trains of pulses. Despite the absence of any
explicit feedback, under suitable conditions we find that the system
displays adaptive behavior including non-associative
learning~\cite{Burrell2001,Rankin2009}, viz., habituation
(desensitization)
and sensitization. These allow plasticity in the behavioral repertoire
of the intracellular signaling motif by enabling modification of the
strength, duration and even the qualitative nature of its response to
recurrent stimulation. In addition to these, we report the occurrence
of a temporal sequence of strong and weak responses to successive
pulses reminiscent of the phenomenon of ``alternans'' in excitable
cells~\cite{Karma2002,Goldhaber2005}. This, coupled with the existence of a response
threshold and an apparent refractory behavior when subjected to
high-frequency stimulation strongly suggests an analogy with excitable
media~\cite{Sridharbook}. While learning is commonly associated with
the behavior at the level of
organisms~\cite{Hennessey1979,Hawkins1984,Thompson1986,Brembs2002,Rankin2006,Lyon2015}, 
it is intriguing that rudimentary forms of such complex adaptive
responses can be seen in a simple network of sub-cellular components.
As the MAPK signaling cascade is involved in coordinating diverse processes
in all eukaryotic cells, these results point to the potential functional
utility of such emergent dynamical phenomena in biological systems.

\begin{figure}
\begin{center}
\includegraphics[width=0.99\linewidth]{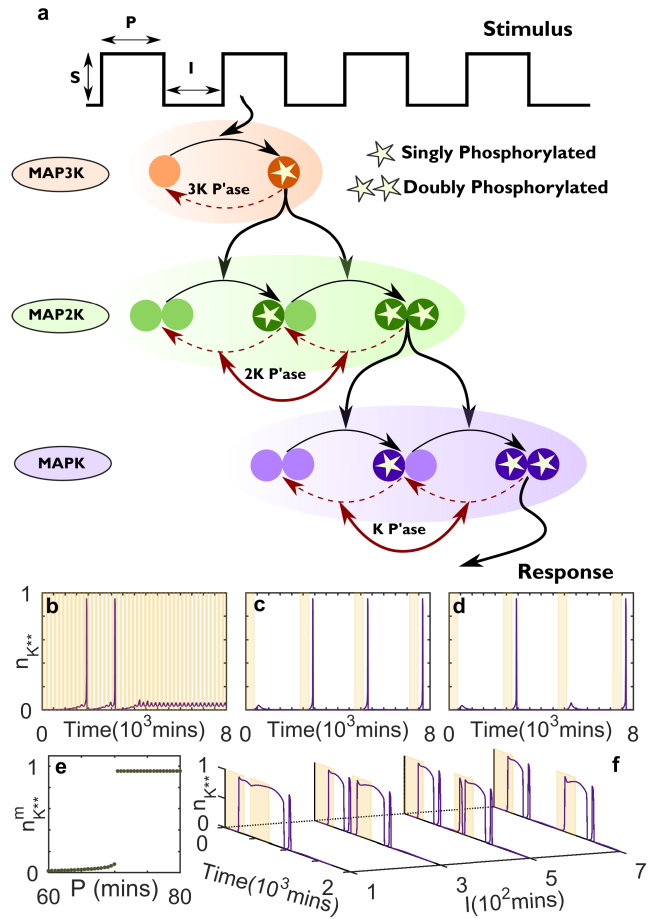}
\end{center}
\caption{Non-associative learning in a MAPK cascade stimulated by a
pulse train. (a) Schematic representation of a linear three-layer MAPK
cascade whose component kinases are activated/deactivated by the
addition/removal of phosphate groups through
phosphorylation/dephosphorylation respectively. Signaling is initiated
when MAPK kinase kinase (MAP3K) is activated by a periodic signal
comprising a series of pulses having amplitude $S$ and duration $P$,
separated by inter-pulse interval $I$. For the cases investigated here,
the cascade receives no stimulus between two successive pulses. The
response of the cascade to the signal is measured in terms of MAPK
activity, viz., the normalized concentration $n_K^{**}$ of doubly
phosphorylated MAPK. (b-d) Time series representing qualitatively
different adaptive responses of the cascade to pulse trains characterized
by a range of $S$, $P$ and $I$. The shaded bars correspond to the
intervals during which MAP3K is stimulated. (b) Desensitization
behavior of the cascade corresponding to an attenuated response on
persistent exposure to the periodic stimulus. (c) Sensitization of the
cascade characterized by a low level of MAPK activity on initial exposure
followed by stronger responses upon repeated stimulation. (d) Alternating
high and low levels of MAPK activity (``alternans'') in response to
successive pulses. (e) Threshold-like response to the pulse duration $P$
of the maximum MAPK activity for a fixed set of values of the signal
strength $S$ and inter-pulse duration $I$ of the pulse train.
(f) Nonlinear dependence of the MAPK cascade response on the inter-pulse
interval for a pair of pulses (shaded bars). 
For details of system and signal parameter values used see
SI.
}
\label{fig:fig1}
\end{figure}

We have simulated the dynamics of the three layer kinase cascade
using the Huang-Ferrell model of the MAPK signaling motif,
schematically illustrated in Fig.~\ref{fig:fig1}~(a). This model consists
of $10$ enzyme-substrate reactions described by $18$ coupled differential
equations~\cite{Ferrell1996}. Each of the several kinase and
phosphatase-mediated enzyme-substrate reactions in the cascade consist
of (i) a reversible enzyme-substrate complex formation step, and (ii) an
irreversible step corresponding to the activation/deactivation of a kinase
(see Supplementary Information for details). The ratio of the activation
and deactivation rates range over four orders of magnitude~\cite{Qiao2007},
underlining the vast diversity of dynamical time-scales present in the
system. The equations are numerically solved without invoking the
quasi-steady-state hypothesis~\cite{Wolkenhauer2007}. We explicitly ensure
that the total concentrations of each of the constituent kinases in the
system are conserved. In our simulations, we assume that the cascade is
initially in the resting state, where the kinases are completely
non-phosphorylated. Following the exposure of the cascade to a train of
pulses, we record the resulting response pattern, viz., the MAPK activity.


Investigations into the dynamics of the Huang-Ferrell
model~\cite{Ferrell1996} have typically focused on the asymptotic response
of the cascade to sustained stimulation. In contrast, here we investigate
the response of the system when it is subjected to recurrent activation by
periodic stimuli. Specifically, we consider a signal comprising a train
of pulses, each having amplitude $S$, duration $P$ and separated from each
other by an inter-pulse interval $I$ [Fig.~\ref{fig:fig1}~(a)].
The cascade is released from stimulation between two successive
pulses, and attempts to relax back to its resting state.
On arrival of the next pulse, the cascade is activated once more,
albeit before it has completely relaxed. This, coupled with the multiple
time-scales of activation and relaxation present in the system, results in
non-trivial adaptive temporal response. 
Selected examples of such behavior are shown
in Fig.~\ref{fig:fig1}~(b-d). These different time series of the activated
MAPK concentration (normalized with respect to the total MAPK concentration) 
correspond to the cascade being subjected to pulse trains characterized by
different parameter values of $P$ and $I$.

Fig.~\ref{fig:fig1}~(b) displays the response of the system subjected
to high-frequency stimulation by short-duration pulses. Here, starting
from its resting state value, each subsequent pulse elicits a slightly
higher response of $n_K^{**}$
until the peak activation suddenly spikes to a value close to
its saturation. This behavior can be interpreted as a form of
signal integration, and may be repeated multiple times as the pulse
train is continued. However, for an
appropriate range of $P$ and $I$ (as in the figure), after a
given number of pulses we observe behavior analogous to {\em
desensitization} when the system no longer shows spiking
activity even for sustained periodic stimulation. 
Thus, following an initial large amplitude response, the subsequent
activity of the system is attenuated even though the nature of the received
signal remained unchanged.
When the cascade is stimulated instead by low-frequency
pulse trains having relatively longer pulse durations, we observe a
phenomenon analogous to {\em sensitization}. Here the cascade exhibits
low-level activity on receiving the initial
pulse but switches to high-amplitude spiking in response to all subsequent
pulses [Fig.~\ref{fig:fig1}~(c)]. 
Thus, the initial low-level activity effectively ``primes'' the
cascade to reach response levels close to saturation. This occurs
because of the existence of long relaxation time-scales in certain
components of the cascade, allowing for response accumulation over
successive stimulations.
Decreasing $P$ by a small amount gives rise to a qualitatively
distinct phenomenon characterized by alternating low and high peaks of
MAPK activity, reminiscent of {\em
alternans}~\cite{Goldhaber2005}.
%

As alternans is a phenomenon that is associated with excitable media,
it is intriguing to consider whether the periodically stimulated
cascade exhibits other characteristics of such systems, in particular,
the existence of a response threshold~\cite{Sridharbook}. As seen in
Fig.~\ref{fig:fig1}~(e), there is indeed a large discontinuous change
in the peak activation $n^{m}_{K^{**}}$ of MAPK when the pulse
duration $P$ crosses a specific value $P_c$ that depends on the choice of
$S$ and $I$. Extending the analogy with excitable media, we find that
the cascade also exhibits a nonlinear relation between its response
to successive pulses and the inter-pulse interval. This can be seen
from the behavior displayed in Fig.~\ref{fig:fig1}~(f) where the
cascade is stimulated by a pair of pulses separated by an interval
$I$. When $I$ is reduced, the response duration resulting from the
second pulse increases in comparison to the duration of the response
caused by the first. As an aside, we note that for the parameter
regime considered here, the system exhibits post-stimulus
reverberatory activity~\cite{Mitra2018}. 

\begin{figure} [hbt!] 
\begin{center}
\includegraphics[width=0.99\linewidth]{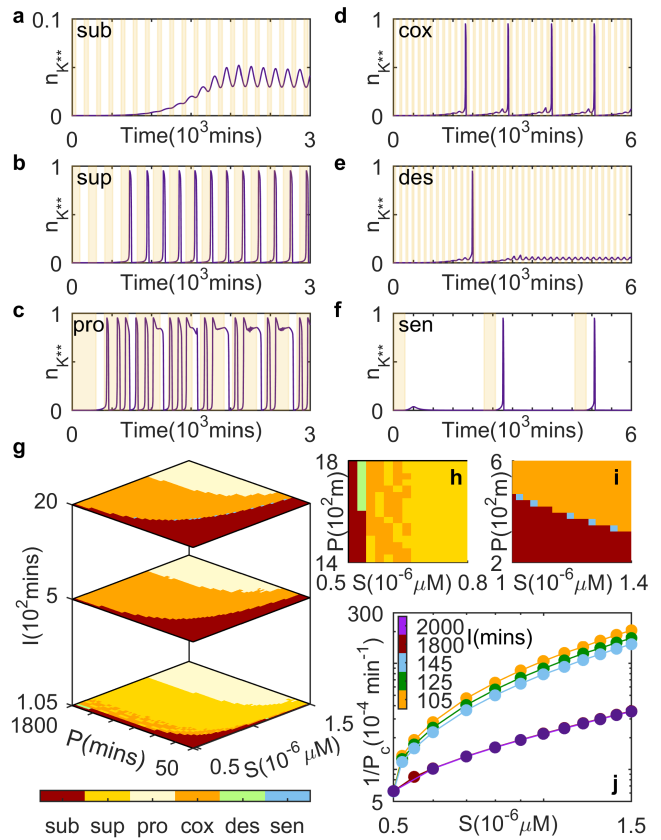}
\end{center}
\caption{(a-f) Characteristic responses of the MAPK cascade to stimulation of
MAP3K by a train of pulses, each of amplitude $S$ having duration $P$,
with inter-pulse interval $I$: (a) attenuated response of the cascade
characterized by sub-threshold activity (sub), (b) large-amplitude
spiking responses characterizing supra-threshold activity (sup), 
(c) prolongation of supra-threshold activity duration (pro) on
application of a signal having pulses with longer duration, (d) coexistence
(cox) of sub- and supra-threshold activity which, for a range of $I$, results
from the integration of responses over several preceding pulses,
(e) desensitization (des), where integration over multiple successive pulses
results in a supra-threshold spiking response but subsequently only exhibits
sub-threshold activity, and (f) sensitization (sen), where sub-threshold
activity in response to the initial pulse gives way to supra-threshold
activity for all subsequent pulses. The shaded bars correspond to the
intervals during which MAP3K is stimulated. (g) Dependence of the cascade
response on the pulse strength $S$ and duration $P$ for three different
values of the inter-pulse interval $I$. The colors represent the nature of
the response [classified into the categories (a-f) mentioned above]. 
(h-i) Magnified views of the $P-S$ planes for (h) $I=105$ and (i) $2000$
minutes show the regions corresponding to desensitization and sensitization,
respectively. (j) The variation of the critical value of pulse duration
$P_c$, above which the cascade exhibits supra-threshold response, with pulse
amplitude $S$. The curves correspond to pulse trains having different
inter-pulse intervals $I$ (as shown in the colorbar).
}
\label{fig:fig2}
\end{figure}

Fig.~\ref{fig:fig2}~(a-f) depicts representative time-series showing the
activity of the cascade on either side of the response threshold, obtained 
for different choices of the periodic stimulation parameters. We note that
in all of the cases shown here, the system shows a gradual build-up of
activity over multiple pulses before reaching asymptotic peak activity
levels. This corresponds to signal integration (mentioned earlier) where
the response of the system to successive stimuli is modulated by the
preceding stimuli. Fig.~\ref{fig:fig2}~(a) shows a typical subthreshold
response ({\em sub}), where the peak MAPK activity is highly attenuated
($< 5\%$ of the saturation response value). Note that the nature of the
response (i.e., whether it is sub- or supra-threshold) is a function of
all three stimulation parameters $S$, $P$ and $I$. For instance, for the
same signal strength $S$ considered in panel (a), the steady-state
response of the cascade would have been close to saturation if the 
stimulation had been applied in a sustained fashion (i.e., $I \rightarrow 0$).

Panels~(b-f) of Fig.~\ref{fig:fig2} depict a variety of suprathreshold
temporal behavior, the simplest of which is characterized by a $1:1$
spiking response to the periodically applied pulses [{\em sup},
Fig.~\ref{fig:fig2}~(b)]. On varying the different stimulation
parameters we observe other types of suprathreshold activity. For
example, on increasing $P$ alone (or alternatively, $S$ alone), the
system exhibits prolongation of the peak activity close to saturation 
[{\em pro}, Fig.~\ref{fig:fig2}~(c)]. For high-frequency stimulation
(i.e., low $I$) after a transient period we observe suprathreshold
peak responses only after every $N$ pulses 
for values of $P$ and $S$ that lie between those giving rise to
{\em sub} and {\em sup} responses [see the lowest plane of
Fig.~\ref{fig:fig2}~(g)]. This response behavior, which corresponds
to the coexistence ({\em cox}) of peak activity levels having
different amplitudes (ranging from values just above zero to
near-saturation) is shown in Fig.~\ref{fig:fig2}~(d). For lower
frequency stimuli, the {\em cox} regime corresponds to $M:1$
response where multiple peaks in MAPK activity, whose amplitudes
can again vary widely, are observed in response to each pulse
[not shown]. Apart from these, we also observe behavior
corresponding to non-associative learning, viz., desensitization
[{\em des}, Fig.~\ref{fig:fig2}~(e)] and sensitization [{\em sen},
Fig.~\ref{fig:fig2}~(f)], as described earlier. Specifically, at
the interface of the {\em cox} and {\em sub} regions in the
stimulation parameter space,
the {\em des} response regime is observed for high-frequency pulse trains
[Fig.~\ref{fig:fig2}~(h)] while for low-frequency
stimulation we obtain {\em sen} [Fig.~\ref{fig:fig2}~(i)].
We note that for higher values of $S$, the transition from {\em cox} to
{\em sub} gets sharper thereby reducing the range of $P$ over which
the {\em des} and {\em sen} phenomena are observed.
An overview of the stimulation parameter space is given in
Fig.~\ref{fig:fig2}~(g) indicating the conditions for which each of
the responses described above can be obtained.

The ``learning'' behavior associated with the periodically stimulated
cascade is seen in the vicinity of the response threshold mentioned
earlier corresponding to the boundary of the {\em sub} regime
[Fig.~\ref{fig:fig2}~(g)]. Hence, we examine the dependence of the
threshold on the stimulation parameters in Fig.~\ref{fig:fig2}~(j).
The reciprocal relation between the signal strength $S$ and
the critical pulse duration $P_c$ necessary for suprathreshold
response seen over a wide range of $S$ suggests that the total signal
intensity of a pulse, measured as the product of $S$ and $P$,
determines the threshold. Deviation from this simple relation is
observed for sufficiently low signal strength. This implies that a minimal
value of $S$ is required to observe suprathreshold response, regardless
of the duration for which the pulse is maintained. We note that in the
limit of $I \rightarrow 0$, this minimal signal strength corresponds
to the lower critical value required to observe a transition from low
level of MAPK activity to high-amplitude oscillations when the cascade
is subjected to sustained stimulation~\cite{Qiao2007,Mitra2018}. As $I$ is
increased, we observe that the response threshold (measured in terms
of the critical pulse duration $P_c$) increases, which suggests that the
excitability of the system reduces as the frequency of the
periodic stimulus decreases.


The phenomena reported here are robust with respect to variations
in the model parameters around the values used in this paper, 
including the kinetic
rate constants and the molecular concentrations of the
constituent kinases and phosphatases. We have also observed similar
behavior with cascades having branched architecture, e.g., MAP3K
activating two different types of MAP2K~\cite{Sitabhra2013}.
While we have assumed that the same phosphatase acts on both the
singly and doubly phosphorylated forms
of the kinase in a particular layer of the cascade (as in the
canonical Huang-Ferrell model), we have explicitly verified that our
results are not sensitively dependent on this.



A mechanistic understanding of the phenomena reported here is made
difficult by the large number of coupled dynamical variables in the
model that operate across different time-scales. This complexity may
be untangled by using the framework of excitable systems. As alluded
to earlier, many of the characteristic features associated with
excitability are present in the system investigated here. These
include the existence of two qualitatively distinct states of
activation separated by a threshold [Fig.~\ref{fig:fig1}~(e)],
nonlinear response to repeated stimulation [Fig.~\ref{fig:fig1}~(f)],
an apparently refractory behavior as seen most prominently during
desensitization [Fig.~\ref{fig:fig1}~(b)] and phenomena analogous to
alternans [Fig.~\ref{fig:fig1}~(d)]. This appealing analogy provides
a means by which a phenomenological understanding of the emergent
behavior of this complex system might be achieved.
We note that the excitability paradigm has been invoked earlier to explain
aspects of cellular activity in the context of antigen recognition by 
T cells~\cite{Grossman1992,Grossman1996}.
Our results show that the emergent dynamics of MAPK cascade, which is
known to mediate immune response~\cite{Dong2002}, provides an explicit
mechanistic basis for such a theoretical framework for explaining the
adaptive response of the immune system to its
microenvironment.

Among the functionally significant dynamical phenomena reported here,
``learning'' is perhaps the most intriguing.
It confers on the system the ability to modify its behavior
in response to information, which is critical for adapting to a changing
environment. The capability to learn often presupposes the existence of a
feedback that
allows bidirectional communication between the components associated
with receiving a signal and those that initiate a corresponding
response~\cite{Staddon2016}. 
In the kinase cascade investigated here, an explicit
feedback is absent as each layer activates the one immediately
downstream. However, an implicit feedback 
results from the inherent features of kinase
activation, viz., sequestration and multi-site
phosphorylation~\cite{Qiao2007,Kholodenko2004,KholodenkoBS2006,Sitabhra2013}.
This can have non-trivial consequences, such as the appearance of
short-term memory, even when the MAPK cascade is subjected to a
single pulse~\cite{Mitra2018}. 

To conclude, in this paper we have shown that a rich repertoire of responses can be
obtained when the system is exposed to a train of pulses. This
results from the implicit feedback, which orchestrates an interplay
between the periodic stimulus and the diverse activation and
relaxation time-scales of the signaling components. In particular,
the system can exhibit sensitization and desensitization, which are
examples of non-associative learning. These may play an important
role in the cell's ability to function in its natural environment,
where it is continually exposed to signals of varying intensity and
duration. This necessitates an ability to respond selectively to
the received stimuli. Such adaptive mechanisms allow the cell to
ignore persistent background stimuli through habituation
(desensitization) but respond strongly to signals to which it has
been primed through earlier exposure (sensitization). Given that a
single linear cascade exhibits such complex adaptive behavior, it
is intriguing to speculate about the potential capabilities
inherent in the coordinated action of multiple subcellular
processes~\cite{IyengarEm1999}. The mechanism through which
learning at the sub-cellular scale can impact adaptive behavior in
an organism at cellular and possibly higher scales remains an
intriguing question.

SNM is supported by the IMSc Complex Systems Project ($12^{\rm th}$ Plan).
The simulations required for this work were done in the High Performance
Computing facility (Nandadevi and Satpura) of The Institute of
Mathematical Sciences, which is partially funded by DST (Grant no.
SR/NM/NS-44/2009).
We thank James Ferrell, Upinder Bhalla, Tharmaraj Jesan, Uddipan Sarma,
Bhaskar Saha, Jose Faro, Vineeta Bal, J. Krishnan, Mukund Thattai,
Marsha Rosner and Pamela Silver for helpful discussions.




\onecolumngrid

\setcounter{figure}{0}
\renewcommand\thefigure{S\arabic{figure}}
\renewcommand\thetable{S\arabic{table}}
\newpage

\vspace{12cm}

\section*{\Large {\bf SUPPLEMENTARY INFORMATION}\\\vspace{0.8cm}for\\\vspace{0.8cm}``Non-associative learning in intra-cellular signaling networks''}

\vspace{1.0cm}

\subsection*{\Large Tanmay Mitra$^{1,2}$, Shakti N. Menon$^{1}$ and Sitabhra Sinha$^{1,2}$}

\vspace{0.5cm}

\subsubsection*{\large $^1$The Institute of Mathematical Sciences, CIT Campus,
Taramani, Chennai 600113, India.\\
$^2$Homi Bhabha National Institute, Anushaktinagar, Mumbai 400094, India.}

\vspace{1cm}



\section{The Model Equations}
\noindent 

\begin{table}[ht]
\caption{Components of the MAPK Cascade} 
\vskip 0.18cm
\centering 
\begin{tabular}{l l l} 
\hline\hline 
Component & Notation & Symbol\\ [0.5ex] 
\hline 
Mitogen-activated Protein Kinase Kinase Kinase & MAP3K & 3K \\ 
Singly Phosphorylated Mitogen-activated Protein Kinase Kinase Kinase & MAP3K* & 3K* \\
Mitogen-activated Protein Kinase Kinase & MAP2K & 2K \\
Singly Phosphorylated Mitogen-activated Protein Kinase Kinase & MAP2K* & 2K* \\
Doubly Phosphorylated Mitogen-activated Protein Kinase Kinase & MAP2K** & 2K** \\
Mitogen-activated Protein Kinase & MAPK & K \\
Singly Phosphorylated Mitogen-activated Protein Kinase & MAPK* & K* \\
Doubly Phosphorylated Mitogen-activated Protein Kinase & MAPK** & K** \\ 
MAP3K-Phosphatase & 3K P'ase & P$_{\rm 3K}$ \\ 
MAP2K-Phosphatase & 2K P'ase & P$_{\rm 2K}$ \\
MAPK-Phosphatase & K P'ase & P$_{\rm K}$ \\ [1ex]
\hline 
\end{tabular}
\end{table}

\vskip 0.5cm

\noindent
\textbf{The three layer MAPK cascade comprises the following
enzyme-substrate reactions:}

\begin{equation*} 
S + 3K\begin{array}{c} {\stackrel{k_{1} }{\longrightarrow} } \\ {\xleftarrow[{k_{-1} }]{} } \end{array}S.3K\stackrel{k_{2} }{\longrightarrow} S + 3K^{*} 
\end{equation*} 

\begin{equation*} 
P_{3K} + 3K^{*}\begin{array}{c} {\stackrel{kp_{1} }{\longrightarrow} } \\ {\xleftarrow[{kp_{-1} }]{} } \end{array}3K^{*}.P_{3K}\stackrel{kp_{2} }{\longrightarrow} P_{3K} + 3K 
\end{equation*} 

\begin{equation*} 
3K^{*} + 2K\begin{array}{c} {\stackrel{k_{3} }{\longrightarrow} } \\ {\xleftarrow[{k_{-3} }]{} } \end{array}3K^{*}.2K\stackrel{k_{4} }{\longrightarrow} 3K^{*} + 2K^{*} 
\end{equation*} 

\begin{equation*} 
P_{2K} + 2K^{*}\begin{array}{c} {\stackrel{kp_{3} }{\longrightarrow} } \\ {\xleftarrow[{kp_{-3} }]{} } \end{array}2K^{*}.P_{2K}\stackrel{kp_{4} }{\longrightarrow} P_{2K} + 2K 
\end{equation*} 

\begin{equation*} 
3K^{*} + 2K^{*}\begin{array}{c} {\stackrel{k_{5} }{\longrightarrow} } \\ {\xleftarrow[{k_{-5} }]{} } \end{array}3K^{*}.2K^{*}\stackrel{k_{6} }{\longrightarrow} 3K^{*} + 2K^{**} 
\end{equation*} 

\begin{equation*} 
P_{2K} + 2K^{**}\begin{array}{c} {\stackrel{kp_{5} }{\longrightarrow} } \\ {\xleftarrow[{kp_{-5} }]{} } \end{array}2K^{**}.P_{2K}\stackrel{kp_{6} }{\longrightarrow} P_{2K} + 2K^{*} 
\end{equation*} 

\begin{equation*} 
2K^{**} + K\begin{array}{c} {\stackrel{k_{7} }{\longrightarrow} } \\ {\xleftarrow[{k_{-7} }]{} } \end{array}2K^{**}.K\stackrel{k_{8} }{\longrightarrow} 2K^{**} + K^{*} 
\end{equation*} 

\begin{equation*} 
P_{K} + K^{*}\begin{array}{c} {\stackrel{kp_{7} }{\longrightarrow} } \\ {\xleftarrow[{kp_{-7} }]{} } \end{array}K^{*}.P_{K}\stackrel{kp_{8} }{\longrightarrow} P_{K} + K 
\end{equation*} 

\begin{equation*} 
2K^{**} + K^{*}\begin{array}{c} {\stackrel{k_{9} }{\longrightarrow} } \\ {\xleftarrow[{k_{-9} }]{} } \end{array}2K^{**}.K^{*}\stackrel{k_{10} }{\longrightarrow} 2K^{**} + K^{**} 
\end{equation*} 

\begin{equation*} 
P_{K} + K^{**}\begin{array}{c} {\stackrel{kp_{9} }{\longrightarrow} } \\ {\xleftarrow[{kp_{-9} }]{} } \end{array}K^{**}.P_{K}\stackrel{kp_{10} }{\longrightarrow} P_{K} + K^{*}
\end{equation*} 


\noindent
\textbf{The above enzyme-substrate reactions can be expressed in terms
of the following coupled ordinary differential equations (ODEs):}


\begin{eqnarray*}
\dt{\FF1}  &=&    \LL1.\FF2 + \LL2.\FF3 - \LL3.\br{S}.\FF1 \,,\\
\dt{\FF2}  &=&    \LL3.\br{S}.\FF1 - (\LL1 + \LL4).\FF2 \,,\\
\dt{\FF3}  &=&    \LL5.\GG1.\FF4  - (\LL2 + \LL6).\FF3\,,\\
\dt{\FF4}  &=&    \LL4.\FF2 + \LL6.\FF3 - \LL5.\GG1.\FF4\\
           & & + (\LL7  + \LL8).\FF6   - \LL9. \FF4.\FF5\\
           & & + (\LL10 + \LL11).\FF9  - \LL12.\FF4.\FF8\,,\\
\dt{\FF5}  &=&    \LL7.\FF6 + \LL19.\FF7 - \LL9.\FF4.\FF5\,,\\
\dt{\FF6}  &=&    \LL9.\FF4.\FF5  - (\LL7 + \LL8).\FF6\,,\\
\dt{\FF7}  &=&    \LL20.\GG2.\FF8 - (\LL19 + \LL21).\FF7\,,\\
\dt{\FF8}  &=&    \LL8.\FF6 + \LL21.\FF7 - \LL20.\GG2.\FF8 \\
           & &  + \LL10.\FF9 - \LL12.\FF4.\FF8 + \LL23.\FF10\,,\\
\end{eqnarray*}

\begin{eqnarray*}
\dt{\FF9}  &=&    \LL12.\FF4.\FF8  - (\LL11 + \LL10).\FF9\,,\\
\dt{\FF10} &=&    \LL22.\GG3.\FF11 - (\LL23 + \LL24).\FF10\,,\\
\dt{\FF11} &=&    \LL11.\FF9 + \LL24.\FF10 - \LL22.\GG3.\FF11\\
           & & + (\LL25 + \LL26).\FF13 - \LL27.\FF11.\FF12\\
           & & + (\LL28 + \LL29).\FF16 - \LL30.\FF11.\FF15\,,\\
\dt{\FF12} &=&    \LL25.\FF13 + \LL31.\FF14 - \LL27.\FF11.\FF12\,,\\
\dt{\FF13} &=&    \LL27.\FF11.\FF12 - (\LL26 + \LL25).\FF13\,,\\
\dt{\FF14} &=&    \LL32.\GG4.\FF15  - (\LL33 + \LL31).\FF14\,,\\
\dt{\FF15} &=&    \LL26.\FF13 + \LL33.\FF14 -\LL32.\GG4.\FF15\\
           & &  + \LL28.\FF16 - \LL30.\FF11.\FF15 + \LL34.\FF17\,,\\
\dt{\FF16} &=&    \LL30.\FF11.\FF15 - (\LL28 + \LL29).\FF16\,,\\
\dt{\FF17} &=&    \LL35.\GG5.\FF18 - (\LL36 + \LL34).\FF17\,,\\
\dt{\FF18} &=&   \LL29.\FF16 + \LL36.\FF17 - \LL35.\GG5.\FF18\,.
\end{eqnarray*}

\noindent where
\begin{align*}
\br{S} &= \br{S}_{\rm tot}  - \FF2\,,\;\;\\
\GG1   &= [P_{3K}] - \FF3\,,\;\; \\
\GG2   &= [P_{2K}] - \FF7\ - \FF10\,,\;\; \\
\GG4   &= [P_{K}] - \FF14\ - \FF17\,.
\end{align*}

\noindent It is explicitly ensured that the total concentrations of
all individual kinases and phosphatases are conserved at all times. 
The concentrations of the different molecular species can vary over
several orders of magnitudes. We have therefore numerically solved the
equations using low relative and absolute tolerances in order to
ensure the accuracy of the resulting time-series.

\newpage

\section{System Parameters}

The numerical values for the reaction rates used in all our simulations are obtained from
Ref.~\cite{Sitabhra2013}, and are listed in Table~\ref{table:Reaction Rates}.
Please note that these values of kinetic rate constants are very close
to that of Huang-Ferrell base values~\cite{Ferrell1996}. 


\begin{table}[ht]
\caption{Reaction Rates}
\vskip 0.15cm
\centering 
\begin{tabular}{l l l l} 
\hline\hline 
Rate constant & Our base value & Huang-Ferrell value & Units \\ [0.5ex] 
\hline 
$k_{1}$ & 1002 & 1000 & $(\mu M.{\rm min})^{-1}$ \\ 
$k_{-1}$ & 150 & 150 & ${\rm min}^{-1}$ \\
$k_{2}$ & 150 & 150 & ${\rm min}^{-1}$  \\
$kp_{1}$ & 1002 & 1000 & $(\mu M.{\rm min})^{-1}$ \\
$kp_{-1}$ & 150 & 150 & ${\rm min}^{-1}$ \\ 
$kp_{2}$ & 150 & 150 & ${\rm min}^{-1}$ \\
$k_{3}$ & 1002 & 1000 & $(\mu M.{\rm min})^{-1}$  \\
$k_{-3}$ & 30 & 150 & ${\rm min}^{-1}$ \\
$k_{4}$ & 30 & 150 & ${\rm min}^{-1}$ \\ 
$kp_{3}$ & 1002 & 1000 & $(\mu M.{\rm min})^{-1}$ \\
$kp_{-3}$ & 150 & 150 & ${\rm min}^{-1}$  \\
$kp_{4}$ & 150 & 150 & ${\rm min}^{-1}$ \\
$k_{5}$ & 1002 & 1000 & $(\mu M.{\rm min})^{-1}$ \\ 
$k_{-5}$ & 30 & 150 & ${\rm min}^{-1}$ \\
$k_{6}$ & 30 & 150 & ${\rm min}^{-1}$  \\
$kp_{5}$ & 1002 & 1000 & $(\mu M.{\rm min})^{-1}$ \\
$kp_{-5}$ & 150 & 150 & ${\rm min}^{-1}$ \\ 
$kp_{6}$ & 150 & 150 & ${\rm min}^{-1}$ \\
$k_{7}$ & 1002 & 1000 & $(\mu M.{\rm min})^{-1}$  \\
$k_{-7}$ & 30 & 150 & ${\rm min}^{-1}$ \\
$k_{8}$ & 30 & 150 & ${\rm min}^{-1}$ \\ 
$kp_{7}$ & 1002 & 1000 & $(\mu M.{\rm min})^{-1}$ \\ 
$kp_{-7}$ & 150 & 150 & ${\rm min}^{-1}$ \\
$kp_{8}$ & 150 & 150 & ${\rm min}^{-1}$  \\
$k_{9}$ & 1002 & 1000 & $(\mu M.{\rm min})^{-1}$ \\
$k_{-9}$ & 150 & 150 & ${\rm min}^{-1}$ \\ 
$k_{10}$ & 150 & 150 & ${\rm min}^{-1}$ \\
$kp_{9}$ & 1002 & 1000 & $(\mu M.{\rm min})^{-1}$  \\
$kp_{-9}$ & 150 & 150 & ${\rm min}^{-1}$  \\
$kp_{10}$ & 150 & 150 & ${\rm min}^{-1}$ \\ [1ex] 
\hline 
\end{tabular}
\label{table:Reaction Rates} 
\end{table}

\newpage

The signal parameters used to generate representative time-series in Fig.~1--2 following the introduction of a signal are listed in Table~\ref{table:System Parameters fig 1} and Table~\ref{table:System Parameters fig 2} respectively.

\begin{table}[ht]
\caption{Signal parameters for the panels in Fig.~1}
\vskip 0.05cm
\centering 
\begin{tabular}{l l l l l l l} 
\hline\hline 
Parameter & (b) & (c) & (d) & (e) & (f) & Units \\ [0.5ex] 
\hline 
$S$ & 1.2 & 1.2 & 1.2 & 1.2 & 3 & $\times 10^{-6} \mu M$ \\
$P$ & 71 & 372 & 371 & 60 -- 80 & 300 & $mins$ \\
$I$ & 106.75 & 2000 & 2000 & 105 & 100 -- 700 & $mins$ \\
\hline 
\end{tabular}
\label{table:System Parameters fig 1} 
\end{table}

\begin{table}[ht]
\caption{Signal parameters for the panels in Fig.~2}
\vskip 0.05cm
\centering 
\begin{tabular}{l l l l l l l l} 
\hline\hline 
Parameter & (a) & (b) & (c) & (d) & (e) & (d) & Units \\ [0.5ex] 
\hline 
$S$ & 1.5 & 1.5 & 1.5  & 1.5 & 1.5 & 1.5 & $\times 10^{-6} \mu M$ \\
$P$ & 49 & 100 & 300 & 50 & 49.5 & 287 & $mins$ \\
$I$ & 105 & 105 & 105 & 105 & 105 & 2000 & $mins$ \\
\hline 
\end{tabular}
\label{table:System Parameters fig 2} 
\end{table}

\begin{table}[ht]
\caption{Total concentration (in $\mu M$) of the kinases and phosphatase proteins for Figs.~1--2} 
\vskip 0.18cm
\centering 
\begin{tabular}{l l} 
\hline\hline 
Protein & Value \\ [0.5ex] 
\hline 
$[K]_{tot}$ & 4.8 \\
$[2K]_{tot}$ & 1.2 \\
$[3K]_{tot}$ & 0.0030 \\
MAP3K-Phosphatase & $1 \times 10^{-4}$  \\ 
MAP2K-Phosphatase & $3 \times 10^{-4}$  \\
MAPK-Phosphatase & 0.05   \\ [1ex]
\hline 
\end{tabular}
\label{table:phosphatases} 
\end{table}

\clearpage

\end{document}